\documentclass[prb,aps,showpacs,superscriptaddress, floatfix]{revtex4}
\usepackage{graphicx}
\usepackage{bm}
\usepackage{amsmath}
\begin{document}
\title{Self-trapping phenomenon in (2+1)-dimensional quantum
electrodynamics and its application to high-T$_c$ cuprates with N{\'
e}el ordering}
\author{T. Morinari}
\affiliation{Yukawa Institute for Theoretical Physics, Kyoto
University Kyoto 606-8502, Japan}
\date{\today}
\begin{abstract}
The strong coupling effect in the (2+1)-dimensional quantum
electrodynamics (QED$_3$) description of the $S=1/2$ Heisenberg
antiferromagnet is studied in terms of a canonical transformation
which has been used in the small polaron theory in electron-phonon
systems.
In the N{\' e}el ordered phase, we show that the Dirac fermions are
localized and the spectral function is of the Gaussian form due to the 
coupling to the longitudinal gauge field.
The width of the broad line shape is $\sim 3J$ with $J$ 
the superexchange interaction energy and the localization length is
$\sim 2a$, with $a$ the lattice constant.
\end{abstract}



\maketitle

\section{Introduction}
One of the most fundamental question in the physics of
high-$T_c$ cuprates is about the description of the doped holes.
The simplest situation would be realized in the single hole doped
system.
Experimentally, angle resolved photoemission spectroscopy (ARPES) in
the undoped compounds provides us valuable information about the
nature of the excitations.
The excitation spectrum was obtained by Wells {\it et
al.}\cite{WELLS_ETAL,RONNING_ETAL,SHEN_RMP}
It was found that the band width is $\sim 2.2J\simeq 270{\rm meV}$ 
where $J$ is the superexchange interaction between the copper site
spins. 
This band width is much smaller than 
the tight-binding value of $8t\simeq 2.8{\rm eV}$ 
which is determined by the transfer 
integral parameter, $t$.
Furthermore, the observed spectra exhibit a quite broad feature.
In Ref.\cite{SHEN_ETAL},  it is shown that the spectra do not have a
conventional Lorentzian form.
The spectra rather have a Gaussian form.
These observations suggest
that the quasiparticle excitations observed in the undoped 
cuprates are quite different from conventional Fermi liquid
quasiparticles.

In order to understand ARPES experiments in the undoped compounds,
first we need to figure out the elementary excitation whose excitation
spectrum obeys the dispersion observed experimentally.
The next step is to consider a mechanism of spectral shape broadening
that would arise from coupling to a boson mode.
Concerning the first point, it was shown that the slave-fermion
approach to the t-t'-t''-J model reproduces the dispersion.
\cite{BOZ95,tttJ}
(It is known that the simple t-J model fails to reproduce 
the dispersion along $(\pi,0)$ to $(0,\pi)$.\cite{TJ_fail,TJ_fail2})
However, it has been shown\cite{BOZ95} that the coupling to
antiferromagnetic spin-wave modes does not lead to a broad line shape.
Recently Mischenko and Nagaosa\cite{MN04} 
considered a coupling to an optical phonon. 
They numerically summed over Feynman diagrams including vertex
corrections for phonons.
It was argued that the quasiparticle in the t-J model is in the strong 
coupling regime that leads to a broad line shape.

In this paper, we consider another possibility.
We consider the quasiparticles in the $\pi$-flux phase
\cite{AFFLECK_MARSTON} and the
effect of coupling to U(1) gauge field fluctuations.
A similar problem within the slave-fermion approach to the t-J model
was discussed by Auerbach and Larson.\cite{AL91}
It was argued that the single hole forms a small polaron due to the
coupling to the gauge field.
Here we consider a similar problem based on the $\pi$-flux phase with
dynamically induced mass.
The $\pi$-flux phase was proposed by Affleck and
Marston\cite{AFFLECK_MARSTON} from a mean 
field theory of the $S=1/2$ antiferromagnetic Heisenberg model
based on a fermionic representation of the spin $S=1/2$.
Fluctuations about the state can be described by a U(1) gauge field.
The dispersion of the quasiparticle in the $\pi$-flux phase is in good 
agreement with the experimentally obtained
dispersion as argued by Laughlin.\cite{LAUGHLIN97}
The effective theory is described by quantum electrodynamics in
(2+1)-dimension, which is called QED$_3$.\cite{MARSTON,KIM_LEE}

Although the $\pi$-flux phase is different from the N{\' e}el ordering
phase, believed to describe the N{\' e}el ordering
state\cite{MARSTON,KIM_LEE} is the state that includes the dynamically
induced mass term due to the coupling to the gauge field.
The excitation spectra near $(\pm \pi/2,\pm \pi/2)$ are given by a
massive Dirac fermion.
Experimentally the Dirac fermion mass is estimated to be $\sim
1.3J$\cite{RONNING_ETALUN} by fitting the experimentally obtained
dispersion with the quasiparticle dispersion in the $\pi$-flux phase
with the mass term.

Meanwhile Franz and Te\u{s}anovi\'{c} proposed \cite{TF01} 
a QED$_3$ theory starting from the d-wave superconducting
state.\cite{BFN99}
The dynamical mass generation, or the spontaneous chiral symmetry
breaking, is also associated with antiferromagnetic
order.\cite{HERBUT}
Aitchison and Mavromatos\cite{AM96} argued that the gauge-fermion
interaction in QED$_3$ leads to non-Fermi liquid behavior.
Rantner and Wen\cite{RW,KHVESHCHENKO} discussed changes of line shapes
between the
normal state and the superconducting state based on a flux phase
obtained in the SU(2) slave-boson approach to the t-J
model\cite{WEN_LEE} with QED$_3$ for
the description of the spinons.

In the N{\' e}el ordered phase, there is strong logarithmic potential
between the Dirac fermions.
In fact, there is no low-lying spin $1/2$ 
excitations \cite{RS90,CHUBUKOV}
because of the confinement associated with this strong
interaction.\cite{QED3_confinement}
Therefore, in the N{\' e}el phase we need to perform a
non-perturbative analysis.

In this paper, we apply to the QED$_3$ a canonical transformation,
which has been extensively studied in the electron-phonon systems,
\cite{MAHAN}
for the effect of the longitudinal gauge field interaction on the
massive Dirac fermion.
We show that the coupling to the longitudinal gauge field leads to the
localization of the Dirac fermions and the spectral function is
approximately given by a Gaussian form.
The main assumption here is that a doped hole excite a quasiparticle
in the $\pi$-flux state.
One way to interpret this is to consider the slave-boson approach to
the t-J model.
Another way is proposed in Ref.\cite{TM05} where a half-skyrmion spin
texture is created by doped holes.
The excitation spectrum of the half-skyrmion spin texture is the same
as that of the $\pi$-flux phase.
The mass term of the quasiparticles in the $\pi$-flux state is the
excitation gap of the half-skyrmion spin texture.
The effective action of the half-skyrmion spin textures is shown to be 
described by QED$_3$ by applying a duality mapping.\cite{TM05}

The rest of the paper is organized as follows.
In Sec.\ref{sec_model}, we describe the QED$_3$ action and the
application of the canonical transformation.
In Sec.\ref{sec_green}, we calculate the retarded Green's function of
the massive Dirac fermion.
Section \ref{sec_conclusion} is devoted to the conclusion.

\section{Model}
\label{sec_model}
In the continuum, the QED$_3$ action for the $S=1/2$ 
quantum Heisenberg antiferromagnet \cite{MARSTON,KIM_LEE} 
is written as
\begin{eqnarray}
S &=& \int d^3 x {\overline \psi} \left( x \right)
\left( {i\gamma ^\mu D_\mu   - mc_{\rm sw}\sigma_3 } \right)
\psi \left( x \right) 
- \frac{1}{4e_A^2}\int d^3 x F_{\mu \nu } F^{\mu \nu }
\nonumber \\
&=& S_0 + S_{\rm int} + S_A,
\label{eq_qed3}
\end{eqnarray}
where $D_{\mu} = \partial_{\mu} + iA_{\mu}$ is the covariant
derivative.
(The derivation of the action in the $\pi$-flux phase is given in
appendix~\ref{sec_qed3}.)
The Dirac fermion fields describe the quasiparticles in the spin
system.
In the absence of the doped holes, the ground state is that all the
negative energy states of the Dirac fermions are occupied.
While the positive energy states are empty.
Upon hole doping, hole states are created in the band of the negative
energy states.
In writing down the action, we have assumed that there is 
the N{\'e}el ordering.
In the presence of the N{\' e}el ordering,
the mass $m$ is nonzero.\cite{MARSTON,KIM_LEE}
The dynamics of the gauge field arises from the fermion polarization.
\cite{KIM_LEE}
Due to the mass term the gauge field action takes the form of the
Maxwell action.
The gauge charge $e_A$ is related to $m$ through 
$e_A^2 = 3\pi m c_{sw}$.\cite{KIM_LEE}

For static particles, the scalar potential $\phi \equiv A_0/e_A$ 
is calculated by the following equation:
\begin{equation}
\nabla^2 \phi ({\bf r})= -e_A \rho ({\bf r}),
\end{equation}
where $\rho ({\bf r})$ represents the charge density.
The potential between a particle and hole is
given by
\begin{equation}
V(r)=-\frac{e_A^2}{2\pi} \ln \frac{r}{r_0}.
\end{equation}
Here $r_0$ is the short distance cutoff scale which is on the order of
the lattice constant.
Note that the potential $V(r)$ has a form of a logarithmically
confining potential.

We consider the effect of the scalar potential on a dynamical hole.
In solid state physics, the Coulomb gauge is often used for the
analysis of the electromagnetic gauge field.
In the Coulomb gauge, the interaction mediated by the scalar potential 
is instantaneous in time.
While in the Lorentz gauge, the retardation effect in 
the interaction is explicit.
For the electromagnetic gauge field, the speed of light $c$ is much
faster than vectors of excitations in the sample.
Therefore, the retardation effect is negligible.
However, for the QED$_3$ theory of (\ref{eq_qed3}), the speed of the
gauge field propagation is $c_{\rm sw}$ that is much smaller than $c$.
The retardation effect can play an important role.
To include this effect, we take the Lorentz gauge.
In the Lorentz gauge,
the action of the gauge field is 
\begin{equation}
S_A = \frac{c_{\rm sw}}{2e_A^2}
\int dt \int d^2 {\bf r}
\sum_{\mu=0,1,2}
\left[ \frac{1}{c_{\rm sw}^2} 
 \left( \partial_t A_{\mu} \right)^2
- \left( \partial_x A_{\mu} \right)^2
- \left( \partial_y A_{\mu} \right)^2 \right].
\end{equation}
The vector potential obeys the following equation,
\begin{equation}
\frac{1}{c_{sw}^2} \partial_t^2 {\bf A} - \partial_x^2 {\bf A}
- \partial_y^2 {\bf A} = -{\bf j}.
\end{equation}
Note that the equation for the longitudinal component of the vector
potential is derived from the equation for the scalar potential 
in the Lorentz gauge.
Therefore, the independent degrees of freedom are 
the transverse component of the vector potential and 
the scalar potential.
The Hamiltonian for the gauge field is 
\begin{equation}
H_A 
= \frac12 \sum_{q,\mu=0,t} 
\omega_q \left( a_{q\mu}^{\dagger} a_{q\mu} 
+ a_{q\mu} a_{q\mu}^{\dagger} \right),
\end{equation}
where $\omega_q=c_{\rm sw} q$ in the continuum and
we have introduced creation and annihilation operators through
$A_{q\mu} = \sqrt{
e_A^2 c_{\rm sw}/(2\omega_q)
}
\left(
 a_{-q,\mu}^{\dagger} + a_{q\mu} 
\right)$.
Here $\mu=t$ denotes the transverse component of the vector potential.

On the square lattice, the action $S_0+S_{\rm int}$
has the following form\cite{KOGUT83}
\begin{eqnarray}
S_0 + S_{\rm int} &=&
\int  dt~ a^2 
\sum_j 
\left\{
{\overline \psi}_j \left( t  \right)
i\gamma^0  
\partial_t
\psi _j \left( t  \right)
-\overline{\psi}_j(t) c_{\rm sw} \gamma^0 A_0(j,t) \psi_j(t)
- mc_{sw}^2 
\overline \psi  _j \left( t  \right)
\sigma_3
\psi _j \left( t  \right) 
\right. \nonumber \\
& & \left.
- \frac{{c_{sw} }}{{2a}}
{\overline \psi}_j \left( t  \right)
i\gamma^1 
\left[ {e^{iaA_1 \left( j+a{\hat e}_1/2,t \right)} 
\psi_{j + a{\hat e}_1 } \left( t  \right) 
- e^{ - iaA_1 \left( j -a{\hat e}_1/2,t\right)} 
\psi_{j - a{\hat e}_1} 
\left( t  \right)} \right]
\right. \nonumber \\ 
& & \left. 
-
\frac{{c_{sw} }}{{2a}}
{\overline \psi}_j \left( t  \right)
i\gamma^2
\left[ {e^{iaA_2 \left( j +a{\hat e}_2/2,t\right)} 
\psi_{j + a{\hat e}_2 } \left( t  \right) 
- e^{ - iaA_2 \left( j-a{\hat e}_2/2,t \right)} 
\psi_{j - a{\hat e}_2} \left( t  \right)} 
\right] 
\right\}.
\label{eq_s0si}
\end{eqnarray}
The above action contains the gauge field interactions arising from
both the transverse part and the longitudinal part of the gauge field.
The transverse component of the vector potential is important for the
calculation of the quasiparticle lifetime.
Because the scattering process associated with the transverse
component is due to emissions or absorptions of the propagating gauge
fields, or ``photons.''
As demonstrated in Refs.\cite{Nagaosa_Lee,Ioffe_Wiegmann},
the coupling to the transverse vector potential leads to a $T$-linear
resistivity law.
However, the interaction of the holes with the transverse vector
potential does not creat a trapping potential, and
is described by a self-energy effect.
Therefore, the spectrum has the form of Lorentzian which is
inconsistent with the ARPES experiments in the undoped compounds.
Here we focus on the self-trapping phenomenon associated with the
scalar potential, and neglect the transverse component of the vector
potential:
\begin{eqnarray}
H_{\rm int}  &=& c_{\rm sw} a^2 
\sum_{j} 
\overline \psi  _j \gamma^0  A_0(j)
\psi _j
\nonumber \\
&=& \frac{c_{\rm sw} a^2}{\sqrt{\Omega}}
\sum_j \sum_{\bf q}
{\psi}_j^{\dagger}\psi_j
{\rm e}^{i{\bf q}\cdot {\bf R}_j }
A_{{\bf q}}.
\end{eqnarray}

The interaction $H_{\rm int}$ contains the terms that contribute to
the infrared (IR) behavior of QED$_3$\cite{IR,SD,KHVESHCHENKO} and the
other terms.
As is well-known in QED$_3$, the dynamical mass generation is
associated with the IR behavior.\cite{IR_mass}
For the analysis, we need a non-perturbative scheme, such as a $1/N$
expansion with $N$ the number of Dirac fermion species.
For discussion of the dynamical mass generation, the Schwinger-Dyson
equation has been studied.\cite{SD}
In the analysis of the Schwinger-Dyson equation, it is necessary 
to figure out the form of the vertex part.
For the IR region, one needs to properly choose the form of the vertex 
part for the discussion of the dynamical mass generation.
While for the other region, the bare vertex form is known to be
sufficient for the analysis of the system.\cite{KN}
Here we are interested in the interaction with the momentum
which is not in the IR regime.
The IR behavior of the system is understood to be associated with the
mass generation.
Therefore, we introduce an IR cut off $q_0$, which is much smaller
than the ultraviolet cut off $\Lambda\sim \pi/a$,
in the following calculations.

We consider a canonical transformation:\cite{MAHAN}
\begin{equation}
\overline{H}=\exp(s) H \exp(-s),
\label{eq_canonical}
\end{equation}
where
\begin{equation}
s=\sum_{q} \frac{M_{q}}{\omega_q}
\left( a^{\dagger}_{-q} - a_{q} \right),
\end{equation}
with $a_q^{\dagger}=a^{\dagger}_{q0}$ and
$a_q=a_{q0}$, and 
\begin{equation}
M_{q}=\frac{a^2}{\sqrt{\Omega}}
\sum_j 
\sqrt{\frac{e_A^2 c_{\rm sw}^3}{2\omega_q}}
{\psi}_j^{\dagger} \psi_j 
\exp (i{\bf q}\cdot {\bf R}_j ).
\end{equation}
The bare vertex is used here as discussed above.
Hereafter the summation in the momentum space concerning the gauge
fields is to be performed for $q_0<|{\bf q}|<\Lambda$.

Using the following result,
\begin{equation}
\left[ s,\psi_j \right]
= -\frac{a^2 e_A}{\sqrt{2\Omega}}
\sum_{{\bf q}}
\left( \frac{c_{\rm sw}}{\omega_q} \right)^{3/2}
{\rm e}^{i{\bf q}\cdot {\bf R}_j }
\left( a_{-q}^{\dagger} -a_{{\bf q}}
\right) \psi_j,
\end{equation}
we find that the fermion field $\psi_j$ is transformed into
\begin{equation}
e^s \psi_j e^{-s}= X_j \psi_j,
\end{equation}
where 
\begin{equation}
X_j = \exp \left[ 
-\frac{a^2 e_A}{\sqrt{2\Omega}} 
\sum_{q} 
\left( \frac{c_{\rm sw}}{\omega_q} \right)^{3/2}
\exp (i{\bf q}\cdot {\bf R}_j )
 \left( a^{\dagger}_{-q} - a_{q} \right)
\right].
\end{equation}
The transformed Hamiltonian reads
\begin{equation}
\overline{H}=\overline{H}_0+\overline{H}_{\rm int},
\end{equation}
\begin{eqnarray}
\overline{H}_0 &=& 
-\frac{c_{\rm sw}a}{2} \sum_{j,\mu=1,2}
\overline{\psi}_j
\gamma^{\mu} 
\left( 
\psi_{j+a{\hat e}_{\mu}}-\psi_{j-a{\hat e}_{\mu}} \right)
-mc_{\rm sw}^2 \sum_j \overline{\psi}_j \sigma_3 \psi_j
\nonumber \\
& & + \frac{1}{2} \sum_{q}
\omega_q \left( a^{\dagger}_q a_q
+ a_q a^{\dagger}_q \right),
\label{eq_Hbar}
\end{eqnarray}
\begin{equation}
\overline{H}_{\rm int} = 
- \sum_{q} \frac{M_{-q}M_{q}}{\omega_q},
\end{equation}
where we have omitted the band narrowing factor 
$X_j^{\dagger} X_{j\pm a{\hat e}_{\mu}}$,
in the hopping term.
(The estimation is similar to that of eqs.(\ref{eq_delta}) and
(\ref{eq_ell}) below.
See Appendix~\ref{sec_DWF}.)
$\overline{H}_{\rm int}$ describes the logarithmic interaction between
the Dirac fermions.
In the following, we consider the effect of the longitudinal
gauge field on the Dirac fermion spectrum given by $\overline{H}_0$.

\section{Calculation of the Retarded Green's function}
\label{sec_green}
The retarded Green's function may be given by
\begin{equation}
G_R \left( {i,j;t} \right) =  - i\theta \left( t \right)\left\langle
{\psi _i \left( t 
\right)\overline \psi  _j \left( 0 \right) + \overline \psi  _j \left
( 0 \right)\psi _i 
\left( t \right)} \right\rangle 
\end{equation}
We separate this quantity into two terms,
\begin{equation}
G_R(i,j;t) = G_1(i,j;t) + G_2(i,j;t),
\end{equation}
where
\begin{equation}
G_1 \left( {i,j;t} \right) =  - i \theta (t)
\left\langle {\psi _i \left( t
\right)\overline \psi  _j
\left( 0 \right)} \right\rangle,
\end{equation}
\begin{equation}
G_2 \left( {i,j;t} \right) =  - i\theta (t)
\left\langle {
\overline \psi  _j \left( 0 \right)
\psi _i \left( t \right)
} \right\rangle.
\end{equation}

In the following calculations, we only consider the Dirac fermions
with the spin up component and residing near $(\pi/2,\pi/2)$ in the
momentum space.
We redefine the Dirac fermion field $\psi (x)$ so that it represents
the spin up Dirac fermion and describes the components near
$(\pi/2,\pi/2)$.
For the $\gamma$ matrices, we may use the reduced form of $\gamma$
matrices:
$\gamma^0\rightarrow \tau_3$,
$\gamma^1\rightarrow i\tau_2$, and $\gamma^2\rightarrow i\tau_1$ with
$\tau_{\mu}$ the Pauli
matrices because we only consider the two components of the Dirac
fermion fields.
(The calculations below are equally applied to the Dirac
fermions with the spin down component and/or residing near
$(-\pi/2,\pi/2)$.)
For $t>0$, by performing the canonical transformation we have
\begin{eqnarray}
iG_1 \left( {i,j;t} \right) 
&=&  \left\langle {\psi _i \left( t \right)\overline \psi  _j 
\left( 0 \right)} \right\rangle \nonumber \\
&=&  \frac{{Tr\,e^{ - \beta {\overline H}_0 } e^{i{\overline H}_0 t}
X_i \psi _i e^{ - i{\overline H}_0 t} \overline \psi  _j X_j^{\dagger}
}}{{Tre^{ - \beta
{\overline H}_0 } }} \nonumber \\
&=&
\left\langle \exp \left[ { 
- \frac{e_A}{{\sqrt {2\Omega }
   }}
\sum\limits_{q} 
{
\frac{1}{{\omega_q^{3/2} }}
\left( {a_{ - q }^\dag  e^{i\omega _q t}  -
   a_{q} e^{ - 
i\omega _q t} } \right)
   \,e^{iq \cdot R_i } 
} } \right] 
\right. \nonumber \\ & & \left. 
\times \frac{1}{\Omega}\sum\limits_{k,k'}
   {e^{ - 
i\gamma _0 \left( {i\sin k_1 \gamma _1  + i\sin k_2 \gamma _2  + m}
\right)t} \psi _k 
\overline \psi  _{k'} e^{ik \cdot R_i } e^{ - ik' \cdot R_j}}
\right. \nonumber \\
   &  &  \times \left. {\exp \left[ \frac{e_A}{
{\sqrt {2\Omega } }}\sum\limits_{q} {\frac{1}{{\omega _q^{3/2}
}}} 
\left( {a_{ - q}^\dag   
- a_{q} } \right) \,e^{iq
\cdot R_j } 
\right]} \right\rangle,
\label{eq_g1}
\end{eqnarray}
where we have used 
\begin{equation}
{\rm e}^{i{\overline H}_0 t}
\psi _k {\rm e}^{-i\overline{H}_0t}
= \exp \left[ i\left( -\tau_1 \sin k_1 + \tau_2 \sin k_2 + \tau_3 m
\right) t
\right]\psi_k.
\end{equation}
(Hereafter we set $a=1$ and $c_{\rm sw}=1$.)
The matrix in the exponential is diagonalized by the following unitary
matrix:
\begin{equation}
 U_k  = \frac{1}{{\sqrt {2\varepsilon _k \left( {\varepsilon _k  + m} 
\right)} }}\left( {\begin{array}{*{20}c}
   {\varepsilon _k  + m} & { \sin k_1  + i\sin k_2 }  \\
   {-\sin k_1  + i\sin k_2 } & {\varepsilon _k  + m}  \\
\end{array}} \right),
\end{equation}
\begin{equation}
 U_k^\dag   = \frac{1}{{\sqrt {2\varepsilon _k \left( {\varepsilon _k  + m} 
\right)} }}\left( {\begin{array}{*{20}c}
   {\varepsilon _k  + m} & {-\sin k_1  - i\sin k_2 }  \\
   { \sin k_1  - i\sin k_2 } & {\varepsilon _k  + m}  \\
\end{array}} \right),
\end{equation}
with $\epsilon_k = \sqrt{\sin^2 k_1 + \sin^2 k_2 + m^2}$,
and then,
\begin{eqnarray}
iG_1 \left( {i,j;t} \right)  
   &=&  \left\langle \exp \left[ { - \frac{e_A}{{\sqrt {2\Omega }
   }}
\sum\limits_{q} 
{\frac{1}{{\omega _q^{3/2} }}
\left( {a_{ - q}^\dag  e^{i\omega _q t}  - a_{q} e^{ - 
i\omega _q t} } \right)
\,e^{iq \cdot R_i } 
} } \right]
 \right. \nonumber \\ 
& & \left. \times
{\exp \left[ { - \frac{e_A}{{\sqrt {2\Omega } }}\sum\limits_{q} 
{\frac{1}{{\omega _q^{3/2} }}\left( {a_{ - q}^\dag   - a_{q} }
\right)
\,e^{iq \cdot R_j }
} } 
\right]} \right\rangle  \nonumber \\
& & \times \frac{1}{\Omega }\sum\limits_k {U_k }
   \left( 
   {\begin{array}{*{20}c}
   {\left[ {1 - f\left( {\varepsilon _k } \right)} \right]
   e^{-i\varepsilon _k t} } & 0  \\
   0 & {e^{i\varepsilon _k t} f\left( {\varepsilon _k } \right)}  \\
    \end{array}} 
   \right)
  U_k^\dag  e^{ik \cdot \left({R_i-R_j} \right)}.
\end{eqnarray}

Using the following formula,
\begin{equation}
e^{Aa^\dag   + Ba} e^{Ca^\dag   + Da}  = e^{\frac{1}{2}\left( {AB +
CD} \right) + BC} e^{\left( {A + C} \right)a^\dag  } e^{\left( {B + D}
\right)a},
\end{equation}
\begin{equation}
\left\langle {e^{Aa^\dag  } e^{Ba} } \right\rangle  =
e^{AB\left\langle {a^\dag  a} \right\rangle },
\end{equation}
we obtain
\begin{eqnarray}
\lefteqn{
\left\langle 
\exp \left[ -\frac{e_A}{\sqrt{2\Omega}}
\sum_q \frac{1}{\omega_q^{3/2}}
\left( a_{-q}^{\dagger} - a_q \right)
e^{iq\cdot R_j} \right]
\exp \left[ \frac{e_A}{\sqrt{2\Omega}}
\sum_q \frac{1}{\omega_q^{3/2}}
\left( a_{-q}^{\dagger}e^{i\omega_q t} 
- a_q e^{-i\omega_q t} \right)
e^{iq\cdot R_j} \right] \right\rangle}
\nonumber \\
&=& \exp \left[ -\frac{e_A^2}{2\Omega}
\sum_q \frac{1}{\omega_q^3} 
\left(1-e^{i\omega_q t} e^{-iq\cdot (R_i-R_j)}
\right) \right]
\exp \left[ -\frac{e_A^2}{\Omega}
\sum_q \frac{1}{\omega_q^3} 
\left(1-\cos \left[ \omega_q t -q\cdot (R_i-R_j)
\right]
\right) n(\omega_q) \right]
\nonumber \\
&=&
\exp \left[ { i\frac{{e_A^2 }}{{2\Omega }}\sum\limits_q
{\frac{1}{{\omega _q^3 }}\sin \left[ {\omega _q t - q \cdot \left
( {R_i  - R_j } \right)} \right]} } \right] \nonumber \\
& & \times
\exp \left[ -\frac{e_A^2}{2\Omega}
\sum_q \frac{1}{\omega_q^3} \coth \frac{\beta \omega_q}{2}
\left(1-\cos \left[ \omega_q t -q\cdot (R_i-R_j)
\right]
\right) \right]
\label{eq_I}
\end{eqnarray}

Since $\epsilon_k > m \simeq J$, and we are interested in 
the temperature region where $\beta J \gg 1$, we may assume
that $f(\epsilon_k) \ll 1$.
Thus, 
\begin{eqnarray}
iG_1 \left( {i,j;t} \right)
& \simeq & \exp \left[ { i\frac{{e_A^2 }}{{2\Omega }}\sum\limits_q
{\frac{1}{{\omega _q^3 }}\sin \left[ {\omega _q t - q \cdot \left
( {R_i  - R_j } \right)} \right]} } \right] \nonumber \\
& & \times \exp \left[ { - \frac{{e_A^2 }}{{2\Omega }}\sum\limits_q
  {\frac{1}{{\omega _q^3 }}\coth \frac{\beta\omega_q}{2}
  \left( {1 - \cos \left[ {\omega _q t - q \cdot \left( {R_i  -
  R_j } \right)} \right]} \right)} } \right] \nonumber \\ 
& & \times \frac{1}{\Omega }\sum\limits_k {e^{i\varepsilon _k t}
  e^{ik \cdot \left( {R_i  - R_j } \right)} \frac{1}{{2\varepsilon _k
  }}\left( {\begin{array}{*{20}c}
   {m + \varepsilon _k } & { - \sin k_1  - i\sin k_2 }  \\
   { - \sin k_1  + i\sin k_2 } & {\varepsilon _k  - m}  \\
\end{array}} \right)} 
\end{eqnarray}
Performing similar calculations, we obtain
\begin{eqnarray}
iG_2 \left( {i,j;t} \right)
&\simeq & \exp \left[ {i\frac{{e_A^2 }}{{2\Omega }}\sum\limits_q
{\frac{1}{{\omega _q^3 }}\sin \left[ {\omega _q t - q \cdot \left
( {R_i  - R_j } \right)} \right]} } \right] \nonumber \\ 
& & \times \exp \left[ { - \frac{{e_A^2 }}{\Omega }\sum\limits_q
{\frac{1}{{\omega _q^3 }} \coth \frac{\beta\omega_q}{2}
\left( {1 - \cos \left[ {\omega _q t - q \cdot \left( {R_i  -
R_j } \right)} \right]} \right)} } \right] \nonumber \\ 
& &  \times \frac{1}{\Omega }\sum\limits_k {\left( {\begin{array}{*{20}c}
   0 & 0  \\
   0 & 1  \\
\end{array}} \right)e^{ - i\varepsilon _k t} e^{ik \cdot \left( {R_i
- R_j } \right)} } 
\label{eq_Gr}
\end{eqnarray}
Finally, we obtain
\begin{eqnarray}
 iG_R \left( {i,j;t} \right) 
&\simeq & \exp \left[ { - \frac{{e_A^2 }}{\Omega
}\sum\limits_q {\frac{1}{{\omega _q^3 }}
\coth \frac{\beta\omega_q}{2}
\left( {1 - \cos \left[ {\omega _q t - q \cdot \left( {R_i  -
R_j } \right)} \right]} \right)} } \right] \nonumber \\ 
& & 
\times \frac{1}{\Omega }\sum\limits_k {\left( {\begin{array}{*{20}c}
   {e^{i\varepsilon _k t} e^{ik \cdot \left( {R_i  - R_j } \right)}
   f\left( {i,j;t} \right)} & 0  \\
   0 & {e^{ - i\varepsilon _k t} e^{ik \cdot \left( {R_i  - R_j }
   \right)} f^* \left( {i,j;t} \right)}  \\
\end{array}} \right)},
\end{eqnarray}
with
\begin{equation}
f(i,j;t)=
\exp \left[ {i\frac{{e_A^2 }}{{2\Omega }}\sum\limits_q
{\frac{1}{{\omega _q^3 }}
\sin \left( \omega_q t \right)
\cos \left[ q\cdot (R_i-R_j) \right]
}} \right],
\end{equation}
where we have assumed that $(m+\epsilon_k)/2\epsilon_k \simeq 1$, 
$(m-\epsilon_k)/2\epsilon_k \simeq 0$, and $(\sin k_1 \pm i \sin
k_2)/2\epsilon_k \simeq 0$.
Because for the small hole concentration, $k/m\ll 1$.
Note that the factor,
\begin{equation}
\frac{1}{\Omega }\sum\limits_k {e^{ \pm i\varepsilon _k t}
e^{ik \cdot \left( {R_i - R_j } \right)} },
\end{equation}
denotes the wave function of the free propagating particle.

\begin{figure}[htbp]
\includegraphics[scale=0.5]{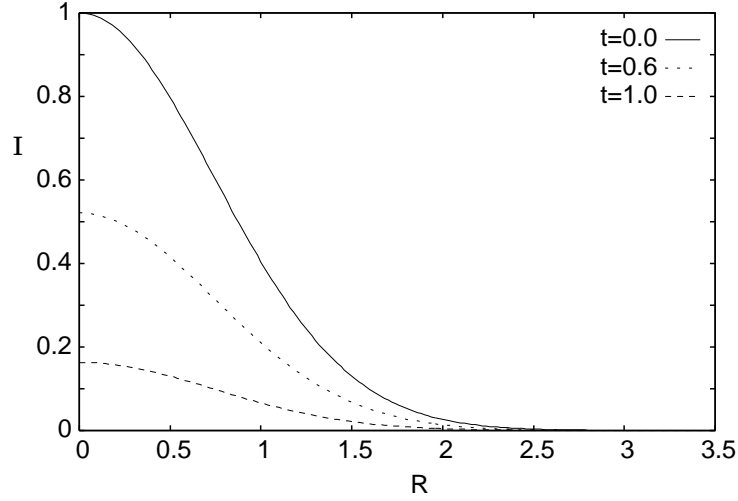}
\caption{The decay factor $I$ as the function of $R=|{\bf
R}_{ij}|$ for several $t$ values. 
The vector ${\bf R}_{ij}$ is assumed to be ${\bf R}_{ij}=(R_{ij},0)$.}
\label{fig_IR}
\end{figure}

\begin{figure}[htbp]
\includegraphics[scale=0.5]{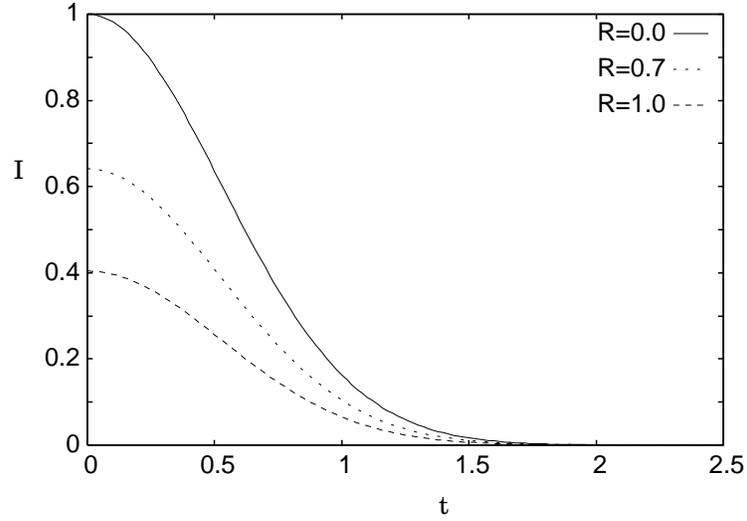}
\caption{The decay factor $I$ as the function of $t$ for
several $R_{ij}(\equiv R)$ values.
The vector ${\bf R}_{ij}$ is assumed to be ${\bf R}_{ij}=(R_{ij},0)$
as in Fig.~\ref{fig_IR}.}
\label{fig_It}
\end{figure}

The second exponential factor ($\equiv I$) in the last line of
Eq.(\ref{eq_I}) leads to a localization effect of the Dirac fermions.
$I$ is approximated as follows,
\begin{equation}
I\simeq -\frac{e_A^2}{8\pi^2} 
\int_0^{\pi} dq \int_0^{2\pi} d\theta
\frac{q}{\omega_q^3} \coth \frac{\beta \omega_q}{2}
\left(1-\cos \left[ \omega_q t -q\cdot (R_i-R_j)
\right]
\right).
\end{equation}
From numerical estimations, we see that $I$ has the Gaussian form as
shown in Figs.~\ref{fig_IR} and \ref{fig_It}.
We find that 
\begin{equation}
I \simeq \exp \left(-1.8125 t^2 -0.90625 R^2\right).
\label{eq_IG}
\end{equation}

To reproduce this Gaussian form,
we make a following approximation:
\begin{eqnarray}
I &\simeq & - \frac{{e_A^2 }}{{4 }} 
\int \frac{d^2 {\bf q}}{(2\pi)^2}
{\frac{1}{{\omega
_q^3 }}\coth \frac{{\beta \omega _q }}{2}\left[ {\omega _q^2 t^2  +
\left( {q \cdot \left( {R_i  - R_j } \right)} \right)^2 } \right]}  
\nonumber \\
&\equiv & - \frac{1}{4}\Delta ^2 t^2  - \frac{{\left( {R_i  - R_j }
\right)^2 }}{{\ell ^2 }},
\end{eqnarray}
where
\begin{equation}
\Delta ^2  = 
e_A^2
\int \frac{d^2 {\bf q}}{(2\pi)^2}
{\frac{1}{{\omega
_q }}\coth \frac{{\beta \omega _q }}{2}} 
\label{eq_delta}
\end{equation}
\begin{equation}
\frac{1}{{\ell ^2 }} = \frac{{e_A^2 }}{{4}}
\int \frac{d^2 {\bf q}}{(2\pi)^2}
{\frac{{q^2 }}{{\omega _q^3 }}\coth \frac{{\beta \omega _q }}{2}} 
\label{eq_ell}
\end{equation}
At low temperatures,
$\Delta$ is
\[
\Delta ^2  \simeq \frac{{e_A^2 }}{\pi }\int_{q_0 }^\Lambda  {dq}
\,\frac{q}{{\omega _q }}\coth \frac{{\beta \omega _q }}{2} =
\frac{{e_A^2 }}{\pi }\left[ {\left( {\Lambda  - q_0 } \right) +
2\sum\limits_{n = 1}^\infty  {\frac{{e^{ - n\beta q_0 }  - e^{ -
n\beta \Lambda } }}{{n\beta }}} } \right] \simeq \frac{{e_A^2 }}{\pi
}\Lambda 
\]
Using the values of $Z_c=1.17$, which
was estimated from quantum Monte Calro simulations,\cite{QMC}
and $m=1.3J$, which was estimated experimentally,
\cite{RONNING_ETALUN}
we obtain
$\Lambda  \simeq \pi /a = \sqrt 2 \pi Z_c J \simeq 5J$,
$e_A^2  = 3\pi m \simeq 12J$.
Thus,
\begin{equation}
\Delta  \simeq 3J.
\end{equation}
This value is in good agreement with the value $\Delta \simeq 2.7J$,
which is obtained from Eq.(\ref{eq_IG}).
Therefore, the above Gaussian approximation is justified.
Similarly $\ell$ is evaluated as
\begin{equation}
\frac{1}{{\ell ^2 }} \simeq \frac{{e_A^2 }}{{8\pi }}\int_{q_0
}^\Lambda  {dq} \coth \frac{{\beta \omega _q }}{2} \simeq \frac{{e_A^2
}}{{8\pi }}\Lambda,
\end{equation}
\begin{equation}
\ell  \simeq 2a.
\end{equation}
From Eq.(\ref{eq_IG}), we find $\ell \simeq a$.
These result suggest that the particle is localized with the
localization length of $\ell = O(a)$.

The function $f(i,j;t)$ is associated with the wave function
renormalization.
The argument of the exponential is
\begin{equation}
\sim \frac{1}{2} Jt\ln \left( \frac{\Lambda}{q_0} \right) \leq \ln
\left( \frac{\Lambda}{q_0} \right).
\end{equation}
(The value of $t$ that gives the saddle point in the frequency
integration may be used in the calculation of $G_R (k,\omega)$ below.)
As a rough approximation, we take $f(i,j;t)$ as a constant $C$.
Thus,
\begin{equation}
iG_R \left( {i,j;t} \right) = \exp \left[ { - \frac{1}{4}\Delta ^2 t^2
- \frac{{\left| {R_i  - R_j } \right|^2 }}{{\ell ^2 }}} \right] \times
\left( {\begin{array}{*{20}c}
   {\frac{C}{\Omega }\sum\limits_k {e^{i\varepsilon _k t} e^{ik \cdot
   \left( {R_i  - R_j } \right)} } } & 0  \\
   0 & {\frac{{C^* }}{\Omega }\sum\limits_k {e^{ - i\varepsilon _k t}
   e^{ik \cdot \left( {R_i  - R_j } \right)} } }  \\
\end{array}} \right).
\end{equation}
Performing Fourier transformation, we obtain
\begin{equation}
iG_R (k,\omega) = \frac{8\pi^{5/2}}{\Delta}
\left( 
\begin{array}{cc}
C\exp \left[
  -(\omega-\epsilon_k)^2/\Delta^2 \right]
& 0 \\
0 & C^* \exp \left[
  -(\omega+\epsilon_k)^2/\Delta^2 \right]
\end{array}
\right)
\end{equation}
The spectral function is 
\begin{equation}
-\frac{1}{\pi} {\rm Im} G_R (k,\omega)
=Z\frac{2\pi^{1/2}}{\Delta}
\left( 
\begin{array}{cc}
\exp \left[
  -(\omega-\epsilon_k)^2/\Delta^2 \right]
& 0 \\
0 & \exp \left[
  -(\omega+\epsilon_k)^2/\Delta^2 \right]
\end{array}
\right),
\end{equation}
where $Z$ is the wave function renormalization factor given by $Z=4\pi
{\rm Re} C$.
Thus, the spectral function has a Gaussian form.
This is consistent with the experiments.\cite{SHEN_ETAL}
The value of the width $\Delta \sim 3J$ appears reasonable compared
with the experiments.
The main drawback in the present calculation is that 
we cannot discuss anisotropy in the momentum space observed in the
experiments.
Because we have started from the continuum theory.
In order to discuss anisotropy, we need to formulate the theory on the 
lattice.

\section{Conclusion}
\label{sec_conclusion}
To conclude, we have studied the effect of the longitudinal gauge
field on the quasiparticle spectrum in the $\pi$-flux phase
with the dynamically induced mass.
A canonical transformation is applied for the analysis of the strong
coupling effect between the massive Dirac fermions and the
longitudinal gauge field.
We have shown that the Dirac fermion propagator has the strong decay
form that leads to the localized state.
The spectral function is approximately given by a Gaussian form.
The width of the broad spectral line shape is $\sim 4J$
and the localization length is $\sim 3a$.

Although we have carried out the analysis in the N{\' e}el ordering
phase, in the disordered phase
we expect a qualitatively different behavior because there is no
strong logarithmic potential that leads to confinement.
It would be interesting to study the delocalization transition
associated with the N{\' e}el order destruction.

\acknowledgments
I would like to thank N.~Nakai and K.~Shizuya for discussions. 
This work was supported by Grant-in-Aid for Young Scientists(B) 
(17740253) and the 21st Century COE "Center for Diversity and
Universality in Physics" from the Ministry of Education, Culture,
Sports, Science and Technology (MEXT) of Japan.

\appendix
\section{Derivation of QED$_3$ action in the $\pi$-flux phase}
\label{sec_qed3}
In this appendix we derive the QED$_3$ action (\ref{eq_qed3}) in the
$\pi$-flux phase\cite{AFFLECK_MARSTON} with the dynamically induced
mass $m$.
The $S=1/2$ antiferromagnetic Heisenberg model is given by
\begin{equation}
H=J\sum_{\langle i,j\rangle} {\bf S}_i \cdot {\bf S}_j,
\end{equation}
where the summation is taken over the nearest neighbor sites.
The spin $1/2$ can be represented by fermion fields,
\begin{equation}
{\bf S}_j = \frac12 
\left( \begin{array}{cc} 
f_{j\uparrow}^{\dagger} &
f_{j\downarrow}^{\dagger}
\end{array} \right)
{\mbox{\boldmath ${\bf \sigma}$}}
\left(
\begin{array}{c}
f_{j\uparrow}\\
f_{j\downarrow}
\end{array}
\right),
\end{equation}
where $f_{j\sigma}$ and $f_{j\sigma}^{\dagger}$ are fermion
annihilation and creation operators at site $j$ with spin $\sigma$,
respectively.
The fermion operators satisfy the constraint,
$\sum_{\sigma} f_{j\sigma}^{\dagger} f_{j\sigma}=1$.
In this fermion representation, the Hamiltonian reads
\begin{equation}
H=-\frac{J}{2} \sum_{\langle i,j\rangle} \sum_{\alpha,\beta}
f_{i\alpha}^{\dagger} f_{j\alpha} f_{j\beta}^{\dagger} f_{i\beta}
+ i\sum_j \lambda_j 
\left( \sum_{\alpha} f_{j\alpha}^{\dagger} f_{j\alpha} -1\right)
+ {\rm const.}
\end{equation}
Here $\lambda_j$ is a Lagrange multiplier to impose the constraint.
Following Affleck and Marston,\cite{AFFLECK_MARSTON} 
we take the mean fields,
\begin{equation}
\chi_{ij} = \sum_{\alpha} 
\langle 
f_{j\alpha}^{\dagger} f_{i\alpha}
\rangle, 
\hspace{2em} \lambda_j = \lambda.
\end{equation}
We set $\chi_1 = \chi_{j,j+{\hat e}_x}$,
$\chi_2 = \chi_{j+{\hat e}_x,j+{\hat e}_y}$,
$\chi_3 = \chi_{j+{\hat e}_x+{\hat e}_y,j+{\hat e}_y}$,
and $\chi_4 = \chi_{j+{\hat e}_y,j}$.
Now the Hamiltonian is
\begin{eqnarray}
H&=& -\frac{J}{2} \sum_{j\in even} \sum_{\sigma}
\left( 
\chi_1 f_{j+{\hat x},\sigma}^{\dagger} f_{j\sigma}
+\chi_4^* f_{j+{\hat y},\sigma}^{\dagger} f_{j\sigma} + h.c.
\right)
-\frac{J}{2} \sum_{j\in odd}
\left( 
\chi_3^* f_{j+{\hat x},\sigma}^{\dagger} f_{j\sigma}
+\chi_2 f_{j+{\hat y},\sigma}^{\dagger} f_{j\sigma} + h.c.
\right) \nonumber \\
& & +\frac{NJ}{4} \left(
|\chi_1|^2+|\chi_2|^2+|\chi_3|^2+|\chi_4|^2
\right)
+ i\lambda\sum_j \left( 
\sum_{\sigma} f_{j\sigma}^{\dagger} f_{j\sigma}-1 \right).
\end{eqnarray}
Performing the Fourier transformation
\begin{equation}
f_{j\sigma} =\frac{1}{\sqrt{N}} \sum_{\bf k}
{\rm e}^{i{\bf k}\cdot {\bf R}_j }
f_{{\bf k}\sigma}
=\frac{1}{\sqrt{N}} {\sum_{\bf k}}^{\prime}
{\rm e}^{i{\bf k}\cdot {\bf R}_j }
\left[
f_{{\bf k}\sigma}
+
(-1)^j f_{{\bf k}+{\bf Q}\sigma} \right],
\end{equation}
with ${\sum_{\bf k}}^{\prime}$ being the summation over the half of
the Brillouin zone,
we obtain
\begin{equation}
H=-\frac{J}{2} {\sum_{\bf k}}^{\prime}
\left( 
\begin{array}{cc}
f_{e{\bf k}}^{\dagger} &f_{o{\bf k}}^{\dagger} 
\end{array} 
\right)
\left( 
\begin{array}{cc}
0 & \kappa_{\bf k}^* \\
\kappa_{\bf k} & 0 
\end{array}
\right)
\left( 
\begin{array}{c}
f_{e{\bf k}} \\
f_{o{\bf k}}
\end{array}
\right) + {\rm const.},
\end{equation}
where the two-component spinors $f_{ek}$ and $f_{ok}$ are
$f_{e{\bf k}} = (f_{\bf k} + f_{{\bf k}+{\bf Q}})/\sqrt{2}$
and $f_{o{\bf k}} = (f_{\bf k} - f_{{\bf k}+{\bf Q}})/\sqrt{2}$,
and $\kappa_{\bf k} = \chi_1 {\rm e}^{-ik_x}+
\chi_3 {\rm e}^{ik_x}+\chi_2^* {\rm e}^{ik_y}+\chi_4^* 
{\rm e}^{-ik_x}$.

In the N{\' e}el ordering phase, there is the staggered magnetization
term
\begin{equation}
\sum_{j} m_{\rm st} (-1)^j f_j^{\dagger} \sigma_3 f_j
= m_{\rm st}{\sum_{\bf k}}^{\prime} 
\left( f_{e{\bf k}}^{\dagger} \sigma_3 f_{e{\bf k}}
-f_{o{\bf k}}^{\dagger} \sigma_3 f_{o{\bf k}}
\right),
\end{equation}
where $\sigma_3$ is a Pauli matrix in spin space.
Following Affleck and Marston,\cite{AFFLECK_MARSTON}
we take
$\chi_1=\chi_3=|\chi|$ and $\chi_2=\chi_4=i|\chi|$. Then,
\begin{equation}
H= {\sum_{\bf k}}^{\prime}
\left( 
\begin{array}{cc}
f_{e{\bf k}}^{\dagger} &f_{o{\bf k}}^{\dagger} 
\end{array} 
\right)
\left( 
\begin{array}{cc}
m_{\rm st} \sigma_3  & J|\chi| \left( \cos k_1 + i\cos k_2 \right) \\
J|\chi| \left( \cos k_1 - i\cos k_2 \right)& -m_{\rm st}\sigma_3
\end{array}
\right)
\left( 
\begin{array}{c}
f_{e{\bf k}} \\
f_{o{\bf k}}
\end{array}
\right) + {\rm const.}
\end{equation}
The low-energy Hamiltonian near $(\pm \pi/2, \pi/2)$ points is
\begin{eqnarray}
H &\simeq &
{\sum_{\bf k}}'
\left( 
\begin{array}{cccc}
f_{e1{\bf k}}^{\dagger} &
f_{o1{\bf k}}^{\dagger} &
f_{e2{\bf k}}^{\dagger} &
f_{o2{\bf k}}^{\dagger}
\end{array} \right)
\left[
\left( \begin{array}{cc} \tau_3 & 0 \\
0 & -\tau_3 \end{array} \right)
m_{\rm st} \sigma_3 
-J|\chi| k_1 
\left( \begin{array}{cc} \tau_1 & 0 \\
0 & -\tau_1 \end{array} \right)
\right. \nonumber \\
& & \left.
+J|\chi| k_2
\left( \begin{array}{cc} \tau_2 & 0 \\
0 & \tau_2 \end{array} \right)
\right]
\left(
\begin{array}{c}
f_{e1{\bf k}}\\
f_{o1{\bf k}}\\
f_{e2{\bf k}}\\
f_{o2{\bf k}}
\end{array}
\right).
\end{eqnarray}
Here $f_{e,o1{\bf k}}$ describes the field near $(\pi/2,\pi/2)$ 
and $f_{e,o2{\bf k}}$ describes the field near $(-\pi/2,\pi/2)$.
The matrices $\tau_j$ are Pauli matrices.
Note that in this representation $(-\pi/2,-\pi/2)$ and
$(\pi/2,-\pi/2)$ are equivalent to 
$(\pi/2,\pi/2)$ and $(-\pi/2,\pi/2)$, respectively.
The Lagrangian is given by
\begin{equation}
L=\int d^2 {\bf r} 
{\psi}^{\dagger} ({\bf r},t)
\left[ i\partial_t
- J|\chi|i\partial_x 
\left(
\begin{array}{cc}
\tau_1 & 0 \\
0 & -\tau_1 
\end{array}
\right)
+J|\chi|i\partial_y 
\left(
\begin{array}{cc}
\tau_2 & 0 \\
0 & \tau_2
\end{array}
\right)
-m_{\rm st}\sigma_3 
\left(
\begin{array}{cc}
\tau_3 & 0 \\
0 & -\tau_3
\end{array}
\right)
\right] \psi ({\bf r},t),
\end{equation}
where ${\psi}^{\dagger} ({\bf r})
= (1/\sqrt{\Omega}) {\sum_k}'\exp \left(- i{\bf k}\cdot {\bf r}
\right) \left( 
\begin{array}{cccc}
f_{e1{\bf k}}^{\dagger} &
f_{o1{\bf k}}^{\dagger} &
f_{e2{\bf k}}^{\dagger} &
f_{o2{\bf k}}^{\dagger}
\end{array} \right)$.
After performing the unitary transformation,
\begin{equation}
\psi ({\bf r},t) \rightarrow 
\left( 
\begin{array}{cc}
1 & 0 \\
0 & i\tau_2 \sigma_1 
\end{array}
\right)\psi ({\bf r},t),
\end{equation}
we obtain
\begin{equation}
L = \int d^2 {\bf r} 
\overline{\psi}({\bf r},t) 
\left[
i\gamma^0 \partial_t -J|\chi| \gamma^1 i\partial_1
-J|\chi| \gamma^2 i\partial_2 - m\sigma_3
\right]\psi ({\bf r},t),
\end{equation}
where $\overline{\psi}=\psi^{\dagger} \gamma^0$,
\begin{equation}
\gamma_0 = \left(
\begin{array}{cc} \tau_3 & 0 \\
0 & -\tau_3 
\end{array}
\right),
\hspace{2em}
\gamma_1 = \left(
\begin{array}{cc} i\tau_2 & 0 \\
0 & -i\tau_2 
\end{array}
\right),
\hspace{2em}
\gamma_2 = \left(
\begin{array}{cc} i\tau_1 & 0 \\
0 & -i\tau_1 
\end{array}
\right).
\end{equation}
Setting $J|\chi| = c_{\rm sw}$,
we obtain
\begin{equation}
S=\int d^3 x \overline{\psi} (x)
\left( i\gamma^{\mu} \partial_{\mu} 
-mc_{\rm sw} \sigma_3 \right) \psi (x),
\end{equation}
where $mc_{\rm sw}^2 = m_{\rm st}$.
Fluctuations about the mean field state are included by a U(1) gauge
field, $A_{\mu}$.\cite{AFFLECK_MARSTON,KIM_LEE}
This is done by replacing $\partial_{\mu}$ with the covariant
derivative
$D_{\mu} = \partial_{\mu} + iA_{\mu}$.

The action of the gauge field $A_{\mu}$ is obtained by calculating the 
fermion polarization function as explicitly shown, for instance,
in Ref.\cite{KIM_LEE}.
Since the fermions are massive, the action has the following
Maxwellian form:
\begin{equation}
S_A= - \frac{1}{4e_A^2}\int d^3 x F_{\mu \nu } F^{\mu \nu },
\end{equation}
The coefficient $1/e_A^2$ is dependent on the mass
through $e_A^2 = 3\pi m$.\cite{KIM_LEE}

We derive this action in the imaginary-time formalism.
If we take the imaginary time as $it=\tau$, 
and define $x_3=c_{sw} \tau$, then
the action takes the so-called Euclidean action:
\begin{equation}
S = \int {d^3 x} \,\overline \psi  \left( x \right)\left[ {\gamma _\mu 
\left( {\partial _\mu   - iA_\mu  } \right) + M} \right]\psi \left( x
\right),
\end{equation}
where $M = mc_{sw}$.
The $\gamma$-matrices are changed as follows,
\begin{equation}
i\gamma ^1  \to \gamma _1 , 
~~i\gamma ^2  \to \gamma _2 , 
~~\gamma ^0 \to \gamma _3.
\end{equation}
Their explicit forms are 
\begin{equation}
\gamma _1  = \left( {\begin{array}{*{20}c}
   { - \tau _2 } & 0  \\
   0 & {\tau _2 }  \\
\end{array}} \right), 
~~\gamma _2  = \left( {\begin{array}{*{20}c}
   { - \tau _1 } & 0  \\
   0 & {\tau _1 }  \\
\end{array}} \right), 
~~\gamma _3  = \left( {\begin{array}{*{20}c}
   {\tau _3 } & 0  \\
   0 & { - \tau _3 }  \\
\end{array}} \right)
\end{equation}
The Dirac fermion Green's function is
\begin{equation}
G_k  =  - \left( {i\gamma _\mu  k_\mu   + M} \right)^{ - 1}.
\end{equation}
One obtains,
\begin{equation}
\left( {ik_3 \tau _3  - ik_1 \tau _2  - ik_2 \tau _1  + M} \right)^{ - 1}  = 
\frac{1}{{M^2  + k^2 }}\left( {\begin{array}{*{20}c}
   { - ik_3  + M} & {k_1  + ik_2 }  \\
   { - k_1  + ik_2 } & {ik_3  + M}  \\
\end{array}} \right)
\end{equation}
Thus, the Green's function is given by
\begin{equation}
G_k  = \frac{1}{{k^2  + M^2 }}\left( {ik_\mu  \gamma _\mu   - M}
\right).
\end{equation}

Now we integrate out the Dirac fermions fields.
After Fourier transformation, the action reads,
\begin{equation}
S = \sum\limits_{k,k'} {\,\overline {\psi _k } \left[ {\left( {ik_\mu
\gamma _\mu + M} \right)\delta _{k,k'}  - i\gamma _\mu  A_\mu  \left
( {k - k'} \right)} \right]} \psi _{k'}.
\end{equation}
Integrating out the Dirac fermions yields
\begin{eqnarray}
S &=&  - Tr\ln \left( { - G_k^{ - 1} } \right) - Tr\ln \left[ {1 +
iG\gamma _\mu  A_\mu  } \right] \nonumber \\
  &=&  - Tr\ln \left( { - G_k^{ - 1}
} \right) - Tr\left[ {iG\gamma _\mu  A_\mu   - \frac{1}{2}\left
( {iG\gamma _\mu  A_\mu  } \right)^2  + ...} \right].
\end{eqnarray}
The effective action of the gauge field is 
\begin{equation}
S^{\left( 2 \right)}  = \frac{1}{2}\int {\frac{{d^3 q}}{{\left( {2\pi
} \right)^3 }}} \Pi _{\mu \nu } \left( q \right)A_\mu  \left( { - q}
\right)A_\nu  \left( q \right).
\end{equation}
The polarization function $\Pi_{\mu \nu} (q)$ is
\begin{equation}
\Pi _{\mu \nu } \left( q \right) =  - \int {\frac{{d^3 k}}{{\left
( {2\pi } \right)^3 }}} tr\left[ {G_k \gamma _\mu  G_{k + q} \gamma
_\nu  } \right]
\end{equation}
Here $tr$ is taken over Dirac fermion components and spin space.

In order to calculate $\Pi_{\mu \nu} (q)$, we use 
the Feynman trick:
\begin{equation}
\frac{1}{{ab}} = \int_0^1 {dx} \,\frac{1}{{\left[ {a\left( {1 - x}
\right) + bx} \right]^2 }}.
\end{equation}
By making use of the Feynman trick, we obtain
\begin{eqnarray}
\Pi _{\mu \nu } \left( q \right) 
&=&  - \int {\frac{{d^3 k}}{{\left( {2\pi } \right)^3 }}} 
\int_0^1 {dx\,\frac{1}{{\left[ {\left( {k +xq} \right)^2  
+ x\left( {1 - x} \right)q^2  + M^2 } \right]^2 }}}
\nonumber \\ 
& & \times tr\left[ {\left( {ik_\mu  \gamma _\mu   - M} \right)\gamma
_\mu  \left( {i\left( {k + q} \right)_\mu  \gamma _\mu   - M}
\right)\gamma _\nu  } \right] \nonumber \\ 
&=&  - \int {\frac{{d^3 k}}{{\left( {2\pi } \right)^3 }}} \int_0^1
  {dx\,\frac{1}{{\left[ {k^2  + x\left( {1 - x} \right)q^2  + M^2 }
  \right]^2 }}}  \nonumber \\   
& & \times tr\left[ {\left( {i\left( {k - xq}
  \right)_\mu  \gamma _\mu   - M} \right)\gamma _\mu  \left( {i\left
  ( {k + \left( {1 - x} \right)q} \right)_\mu  \gamma _\mu   - M}
  \right)\gamma _\nu  } \right].
\end{eqnarray}
To calculate the terms including $\gamma$-matrices,
we use the following formula:
\begin{equation}
\gamma _\mu  \gamma _\nu   = \delta _{\mu \nu }  + i\varepsilon _{\mu
\nu \lambda } \tau _\lambda  \left( {\begin{array}{*{20}c}
   1 & 0  \\
   0 & 1  \\
\end{array}} \right),
\end{equation}
\begin{equation}
tr\left( {\gamma _\mu  \gamma _\nu  } \right) = \delta _{\mu \nu }
tr1,
\end{equation}
\begin{equation}
tr\left( {\gamma _\alpha  \gamma _\mu  \gamma _\beta  \gamma _\nu  }
\right) = \left( {\delta _{\alpha \mu } \delta _{\beta \nu }  - \delta
_{\alpha \beta } \delta _{\mu \nu }  + \delta _{\alpha \nu } \delta
_{\mu \beta } } \right)tr1.
\end{equation}
Thus, we obtain
\begin{eqnarray}
\Pi _{\mu \nu } \left( q \right) 
&=&  - 8\int {\frac{{d^3 k}}{{\left
 ( {2\pi } \right)^3 }}} \int_0^1 {dx\,\frac{1}{{\left[ {k^2  +
 x\left( {1 - x} \right)q^2  + M^2 } \right]^2 }}} \nonumber \\ 
& &  \times \left[ { - 2k_\mu  k_\nu   + 2x\left( {1 - x}
 \right)\left( {q_\mu  q_\nu   - q^2 \delta _{\mu \nu } } \right)}
 \right.\left. { + \left[ {k^2  + x\left( {1 - x} \right)q^2  + M^2 }
 \right]\delta _{\mu \nu } } \right],
\end{eqnarray}
where we have used $tr 1=8$.

As a convenient calculation scheme, we use 
the dimensional regularization:\cite{dim_reg}
\begin{eqnarray}
 \Pi _{\mu \nu } \left( q \right) 
&=&  - \frac{{16\pi ^{D/2} }}{{\left
 ( {2\pi } \right)^D \Gamma \left( {D/2} \right)}}\int_0^1
 {dx\,\int_0^\infty  {dk} \,k^{D - 1} \frac{{2x\left( {1 - x}
 \right)\left( {q_\mu  q_\nu   - q^2 \delta _{\mu \nu } }
 \right)}}{{\left[ {k^2  + x\left( {1 - x} \right)q^2  + M^2 }
 \right]^2 }}} \nonumber  \\ 
&  &  - 8\int {\frac{{d^D k}}{{\left( {2\pi } \right)^D }}} \int_0^1 
{dx\,\frac{{\left[ {k^2  + x\left( {1 - x} \right)q^2  + M^2 }
 \right]\delta _{\mu \nu }  - 2k_\mu  k_\nu  }}{{\left[ {k^2  +
 x\left( {1 - x} \right)q^2  + M^2 } 
\right]^2 }}}.  
\end{eqnarray}
The integrations in the first term is calculated as follows
\begin{eqnarray}
\lefteqn{\int_0^1 {dx\,x\left( {1 - x} \right)\,\int_0^\infty  {dk}
 \,\frac{{k^{D - 1} }}{{\left[ {k^2  + x\left( {1 - x} \right)q^2  +
 M^2 } \right]^2 }}}} \nonumber \\   
&=& \frac{\pi }{2}\frac{{1 - D/2}}{{\sin
 \left( {\pi D/2} \right)}}\int_0^1 {dx} 
\,x\left( {1 - x} \right)\left[ {M^2  + x\left( {1 - x} \right)q^2 }
\right]^{D/2 - 2} 
\end{eqnarray}
Setting $D = 3$, we obtain
\begin{equation}
\int_0^1 {dx} \,x\left( {1 - x} \right)\left[ {M^2  + x\left( {1 - x}
\right)q^2 } \right]^{ - 1/2}  = \frac{M}{{2q^2 }} + \frac{{q^2  -
4M^2 }}{{4q^3 }}\sin ^{ - 1} \frac{q}{{\sqrt {q^2  + 4M^2 } }}.
\end{equation}

The second term is rewritten as
\begin{eqnarray}
\lefteqn{\int {\frac{{d^D k}}{{\left( {2\pi } \right)^D }}} \int_0^1
{dx\,\frac{{\left[ {k^2  + x\left( {1 - x} \right)q^2  + M^2 }
\right]\delta _{\mu \nu }  - 2k_\mu  k_\nu  }}{{\left[ {k^2  + x\left
( {1 - x} \right)q^2  + M^2 } \right]^2 }}}}  \nonumber \\
&=& \delta _{\mu \nu }
\int {\frac{{d^D k}}{{\left( {2\pi } \right)^D }}} \int_0^1
{dx\,\frac{1}{{k^2  + x\left( {1 - x} \right)q^2  + M^2 }}}  - 2\int
{\frac{{d^D k}}{{\left( {2\pi } \right)^D }}} \int_0^1
{dx\,\frac{{k_\mu  k_\nu  }}{{\left[ {k^2  + x\left( {1 - x}
\right)q^2  + M^2 } \right]^2 }}}.
\label{eq_2nd}
\end{eqnarray}
The integral over $k$ in the second term in the right hand side is
carried out as follows
\begin{eqnarray}
\lefteqn{ \int {\frac{{d^D k}}{{\left( {2\pi } \right)^D }}}
\frac{{k_\mu k_\nu  }}{{\left[ {k^2  + x\left( {1 - x} \right)q^2  +
M^2 } \right]^2 }} } \nonumber \\   
&=& \frac{{2\pi ^{D/2} }}{{\left( {2\pi } \right)^D
 D\Gamma \left( {D/2} \right)}}\delta _{\mu \nu } \int_0^\infty  {dk}
 \,\frac{{k^{D + 1} }}{{\left[ {k^2  + x\left( {1 - x} \right)q^2  +
 M^2 } \right]^2 }} \nonumber \\
&=& \frac{{\pi ^{D/2 + 1} }}{{2\left( {2\pi }
 \right)^D \Gamma \left( {D/2} \right)}}\frac{{\left[ {x\left( {1 - x}
 \right)q^2  + M^2 } \right]^{D/2 - 1} }}{{\sin \left( {\pi D/2}
 \right)}}\delta _{\mu \nu }.
\end{eqnarray}
The first term in the right hand side is,
\begin{eqnarray}
\lefteqn{ \int {\frac{{d^D k}}{{\left( {2\pi } \right)^D }}}
\frac{1}{{k^2  + x\left( {1 - x} \right)q^2  + M^2 }} }
\nonumber \\ 
&=& \frac{{2\pi ^{D/2}
 }}{{\left( {2\pi } \right)^D \Gamma \left( {D/2}
 \right)}}\int_0^\infty  {dk} \,\frac{{\,k^{D - 1} }}{{k^2  + x\left
 ( {1 - x} \right)q^2  + M^2 }} \nonumber \\   
&=& \frac{{\pi ^{D/2 + 1}
 }}{{\left( {2\pi } \right)^D \Gamma \left( {D/2}
 \right)}}\frac{{\left[ {x\left( {1 - x} \right)q^2  + M^2 }
 \right]^{D/2 - 1} }}{{\sin \left( {\pi D/2} \right)}}. 
\end{eqnarray}
From these equations, we find that the two terms in the right hand
side of Eq.~(\ref{eq_2nd}) exactly cancel each other.

As a result, we obtain
\begin{equation}
\Pi _{\mu \nu } \left( q \right) = \frac{2}{\pi }\left( {q^2 \delta
_{\mu \nu }  - q_\mu  q_\nu  } \right)\left[ {\frac{M}{{2q^2 }} +
\frac{{q^2  - 4M^2 }}{{4q^3 }}\sin ^{ - 1} \frac{q}{{\sqrt {q^2  +
4M^2 } }}} \right].
\end{equation}
Expaiding the right hand side with respect to $q$, 
we obtain
\begin{equation}
\Pi _{\mu \nu } \left( q \right) = \frac{1}{{3\pi M}}\left( {q^2
\delta _{\mu \nu }  - q_\mu  q_\nu  } \right)\left[ {1 -
\frac{1}{{10}}\left( {\frac{q}{M}} \right)^2  + \frac{1}{{560}}\left
( {\frac{q}{M}} \right)^4  + ...} \right].
\end{equation}

On the other hand, the Maxwell term has the following form
\begin{equation}
S_M  = \frac{1}{{2e_A^2 }}\sum\limits_q {\left( {q^2 \delta _{\mu \nu
}  - q_\mu  q_\nu  } \right)A_\mu  \left( { - q} \right)A_\nu  \left
( q \right)} 
\end{equation}
Thus, we obtain $e^2  = 3\pi M$.

\section{The band narrowing factor}
\label{sec_DWF}
In this appendix, we evaluate the band narrowing factor omitted in
eq.(\ref{eq_Hbar}).
The canonical transformation (\ref{eq_canonical}) introduces the
following factor in the hopping term,
\begin{equation}
X_i^\dag  X_j  = \exp \left[ {\frac{{e_A }}{{\sqrt {2\Omega }
}}\sum\limits_q {\frac{1}{{\omega _q^3 }}\left( {a_{ - q}^\dag   - a_q
} \right)e^{iq \cdot R_i } } } \right]\exp \left[ { - \frac{{e_A
}}{{\sqrt {2\Omega } }}\sum\limits_q {\frac{1}{{\omega _q^3 }}\left
( {a_{ - q}^\dag   - a_q } \right)e^{iq \cdot R_j } } } \right].
\end{equation}
By taking the thermal average with respect to the bosons, we obtain
\begin{equation}
\left\langle {X_i^\dag  X_j } \right\rangle  = \exp \left[ { -
\frac{{e_A^2 }}{{2\Omega }}\sum\limits_q {\frac{1}{{\omega _q^3
}}\coth \left( {\frac{{\beta \omega _q }}{2}} \right)\left( {1 - \cos
\left[ {q \cdot \left( {R_i  - R_j } \right)} \right]} \right)} }
\right].
\end{equation}
The calculation is similar to eqs.(\ref{eq_delta}) and (\ref{eq_ell}).
Thus,
\begin{equation}
\left\langle {X_i^\dag  X_j } \right\rangle  \simeq 0.7
\end{equation}
The hopping amplitude is somewhat reduced by the band narrowing
factor.


\begin{thebibliography}{99} 
\bibitem{WELLS_ETAL}B.~O.~Wells, Z.-X.~Shen, 
A.~Matsuura, D.~M.~King, M.~A.~Kastner, M.~Greven, 
and R.~J.~Birgeneau,
Phys. Rev. Lett. {\bf 74}, 964 (1995).
\bibitem{RONNING_ETAL}
F.~Ronning, C.~Kim, D.~L.~Feng, D.~S.~Marshall, A.~G.~Loeser, 
L.~L.~Miller, J.~N.~Eckstein, I.~Bozovic, and Z.-X.~Shen,
Science {\bf 282}, 2067 (1998).
\bibitem{SHEN_RMP} See, for a review, A.~Damascelli, Z.~Hussain, and
Z.~X.~Shen, Rev. Mod. Phys. {\bf 75}, 473 (2003).
\bibitem{SHEN_ETAL}K.M. Shen, F. Ronning, D.H. Lu, W.S. Lee,
N.J.C. Ingle, W. Meevasana, F. Baumberger, A. Damascelli,
N.P. Armitage, L.L. Miller, Y. Kohsaka, M. Azuma, M. Takano,
H. Takagi, and Z.-X. Shen, Phys. Rev. Lett. {\bf 93}, 267002 (2004).
\bibitem{BOZ95}
J.~Ba{\l}a, A.~M.~Oles, and J.~Zaanen, Phys. Rev. B {\bf 52}, 4597
(1995).
\bibitem{tttJ}
T.~Xiang and J.~M.~Wheatley, Phys. Rev. B {\bf 54}, R12 653
(1996);B.~Kyung and R.~A.~Ferrell, Phys. Rev. B {\bf 54}, 10125
(1996);T.~K.~Lee and C.~T.~Shih, Phys. Rev. B {\bf 55}, 5983 (1997);
T.~K.~Lee, C.-M.~Ho, and N.~Nagaosa, Phys. Rev. Lett. {\bf 90}, 067001
(2003); T.~Tohyama and S.~Maekawa, Supercond. Sci. Technol. {\bf 
13}, R17 (2000).
\bibitem{TJ_fail} Z.~Liu and E.~Manousakis, 
Phys. Rev. B 45, 2425 (1992).
\bibitem{TJ_fail2}
E.~Dagotto, Rev. Mod. Phys. 66, 763 (1994).
\bibitem{MN04}A.~S.~Mishchenko and N.~Nagaosa, Phys. Rev. Lett. {\bf
93}, 036402 (2004).
\bibitem{AFFLECK_MARSTON}I.~Affleck and J.~B.~Marston, Phys. Rev. B 
{\bf 37}, R3774 (1988); J.~B.~Marston and I.~Affleck, Phys. Rev. B
{\bf 39}, 11538 (1989).
\bibitem{AL91}A.~Auerbach and B.~E.~Larson, Phys. Rev. Lett. {\bf 66}, 
2262 (1991).
\bibitem{LAUGHLIN97}R.~B.~Laughlin, Phys. Rev. Lett. 79, 1726 (1997).
\bibitem{MARSTON}J.~B.~Marston, Phys.~Rev.~Lett. {\bf 64}, 1166
(1990).
\bibitem{KIM_LEE}D.~H.~Kim and P.~A.~Lee, Ann. Phys. {\bf 272}, 130
(1999).
\bibitem{RONNING_ETALUN}F.~Ronning, C.~Kim, Z.-X.~Shen, and
L.~L.~Miller, unpublished.
\bibitem{TF01}M~Franz and Z.~Te\u{s}anovi\'{c}, Phys. Rev. Lett. {\bf
87}, 257003 (2001).
\bibitem{BFN99}L.~Balents, M.~P.~A.~Fisher, and C.~Nayak, Phys. Rev. B
{\bf 60}, 1654 (1999).
\bibitem{HERBUT}
Z.~Te\u{s}anovi\'{c}, O.~Vafek and M.~Franz, 
Phys. Rev. B {\bf 65}, 180511(R) (2002);
I.~F.~Herbut, Phys. Rev. Lett. {\bf 88}, 047006 (2002).
\bibitem{AM96}I.~J.~R.~Aitchison and N.~E.~Mavromatos, Phys. Rev. B
{\bf 53}, 9321 (1996) and references cited therein.
\bibitem{RW}W.~Rantner and X.-G.~Wen, Phys. Rev. Lett. {\bf 86}, 3871
(2001);Phys. Rev. B {\bf 66}, 144501 (2002).
\bibitem{KHVESHCHENKO}See, also, D.~V.~Khveshchenko, Phys. Rev. B {\bf
65}, 235111 (2002).
\bibitem{WEN_LEE} X.-G.~Wen and P.~A.~Lee, Phys. Rev. Lett. {\bf 76},
503 (1996).
\bibitem{RS90}N.~Read and S.~Sachdev, Phys. Rev. B {\bf 42},
4568 (1990).
\bibitem{CHUBUKOV}A.~Chubukov, Phys.~Rev.~B {\bf 44}, 12318 (1991).
\bibitem{QED3_confinement}C.~J.~Burden, J.~Praschifka, and
C.~D.~Roberts, Phys. Rev. D {\bf 46}, 2695 (1992);
P.~Maris, Phys. Rev. D {\bf 52}, 6087 (1995);
G.~Grignani, G.~Semenoff and P.~Sodano, Phys.~Rev. D {\bf 53}, 7157 
(1996).
\bibitem{MAHAN}G.~D.~Mahan, {\it Many-Particle Physics} (Plenum, New
York, 1990), 2nd edition.
\bibitem{TM05}T.~Morinari, cond-mat/0507666, to be published in
Phys. Rev. B; cond-mat/0502437.
\bibitem{KOGUT83} J.~B.~Kogut, Rev. Mod. Phys. {\bf 55}, 775 (1983).
\bibitem{Nagaosa_Lee} N.~Nagaosa and P.~A.~Lee, 
Phys. Rev. Lett. {\bf 64}, 2450 (1990).
\bibitem{Ioffe_Wiegmann} L.~B.~Ioffe and P.~B.~Wiegmann, 
Phys. Rev. Lett. {\bf 65}, 653 (1990).
\bibitem{IR}R.~Jackiw and S.~Templeton, Phys. Rev. D {\bf 23}, 2291
(1981).
\bibitem{SD}
T.~W.~Appelquist, M.~Bowick, D.~Karabali and
L.~C.~R.~Wijewardhana, Phys. Rev. D {\bf 33}, 3704 (1986);
\bibitem{IR_mass}T.~Appelquist, D.~Nash and
L.~C.~R.~Wijewardhana, Phys.~Rev.~Lett. {\bf 60}, 2575 (1988).
\bibitem{KN}K.~Kondo and H.~Nakatani, Mod. Phys. Lett. A{\bf 4}, 2155
(1989).
\bibitem{QMC}B.~B.~Beard, R.~J.~Birgeneau, M.~Greven, and U.~J.~Wiese, 
Phys. Rev. Lett. {\bf 80}, 1742 (1998);
J.-K.~Kim and M.~Troyer, Phys. Rev. Lett. {\bf 80}, 2705 (1998).
\bibitem{HH69}B.~I.~Halperin and P.~C.~Hohenberg, Phys. Rev. {\bf
188}, 898 (1969).
\bibitem{CHN} S.~Chakravarty, B.~I.~Halperin, and D.~R.~Nelson,
Phys.~Rev.~B {\bf 39}, 2344 (1989).
\bibitem{HN93}P.~Hasenfratz and F.~Niedermayer, Z. Phys. B 92, 91
(1993).
\bibitem{dim_reg} See, for example, P. Ramond, 
{\it Field Theory: A Modern Primer}
(Addison-Wesley, 1989).
\end{thebibliography}
\end{document}